\documentclass[twocolumn,superscriptaddress,showpacs,prl]{revtex4}    
\usepackage{graphicx,amsmath,mathrsfs}
\usepackage{amssymb}
\usepackage{amsthm,multirow}
\usepackage{bm,bbm}
\usepackage{subfigure,xcolor}   

\usepackage{color}

\def\bc{\begin{center}}
\def\ec{\end{center}}

\newcommand{\bs}[1]{\boldsymbol{#1}}

\newcommand{\pd}{{\phantom{\dagger}}}

\newcommand\dhat{{\hat{\bs{\delta}}}}

\def\ie{\emph{i.e.},\ }
\def\eg{\emph{e.g.}\ }

\begin{document}
\title{
Strain-induced Landau Levels in arbitrary dimensions with an exact spectrum
}
\author{Stephan Rachel}
\affiliation{Institut f\"ur Theoretische Physik, Technische Universit\"at Dresden,
01062 Dresden, Germany}
\author{Ilja G\"othel}
\affiliation{Institut f\"ur Theoretische Physik, Technische Universit\"at Dresden,
01062 Dresden, Germany}
\author{Daniel P.\ Arovas}
\affiliation{Department of Physics, University of California, San Diego, La Jolla, California 92093, USA}
\author{Matthias Vojta}
\affiliation{Institut f\"ur Theoretische Physik, Technische Universit\"at Dresden,
01062 Dresden, Germany}

\begin{abstract}
Certain non-uniform strain applied to graphene flakes has been shown to induce pseudo-Landau levels in the single-particle spectrum, which can be rationalized in terms of a pseudo-magnetic field for electrons near the Dirac points. However, this Landau level structure is in general approximate and restricted to low energies.
Here we introduce a family of strained bipartite tight-binding models in arbitrary spatial dimension $d$ and analytically prove that their entire spectrum consists of perfectly degenerate pseudo-Landau levels. This construction generalizes the case of triaxial strain on graphene's honeycomb lattice to arbitrary $d$; in $d=3$ our model corresponds to tetraxial strain on the diamond lattice.
We discuss general aspects of pseudo-Landau levels in arbitrary $d$.
\end{abstract}


\date{\today}

\maketitle


\begin{figure*}[t!]
\centering
\includegraphics[scale=0.6]{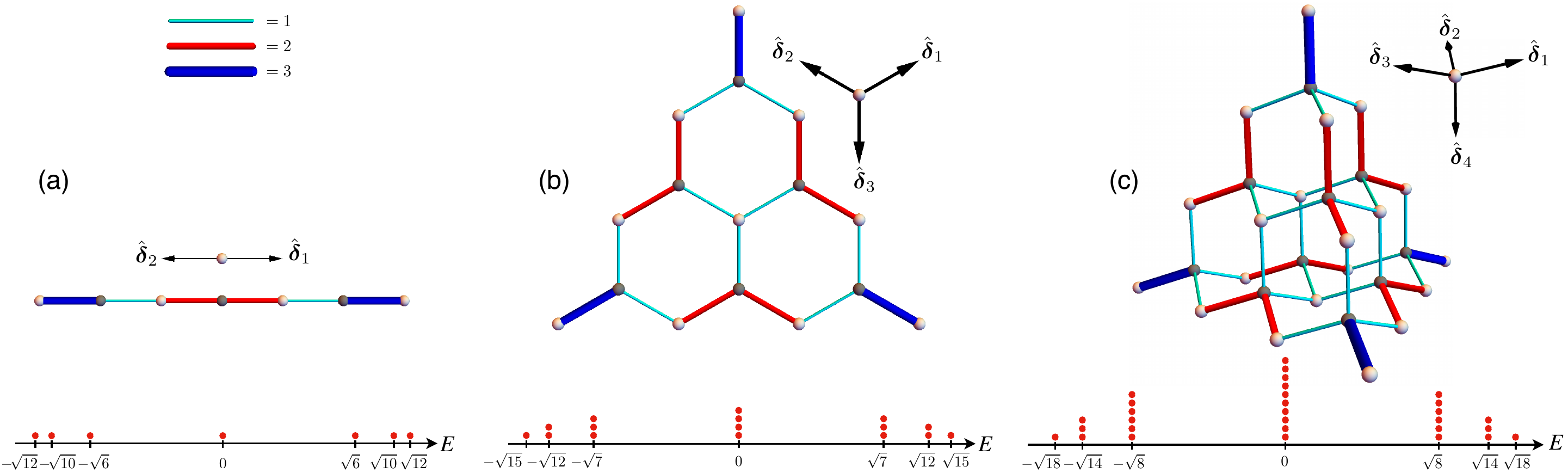}
\caption{
Finite-size lattices (top) and their spectra (bottom) for $N=4$ and (a) $d=1$, (b) $d=2$, and (c) $d=3$.
The hopping amplitudes $t^{(N)}_{\bs r,j}$, Eq.~\eqref{ddim-strain}, are indicated via line thickness and color, see inset. The next outer bonds have vanishing amplitude, thus cutting out (a) a line segment, (b) a triangle, and (c) a tetrahedron, each of linear size $N$. Note that unstrained lattices are shown; for finite electron-lattice coupling weak (strong) bonds would be elongated (compressed).
The energy levels and their degeneracies are given in Eqs.~\eqref{ddim-spec} and \eqref{ddim-deg}, respectively, for details see text.
}
\label{fig:1-2-3}
\end{figure*}

\paragraph{Introduction --}
The engineering of quantum phases and their properties has become an important concept in condensed matter physics. In this approach, one either builds or influences a large quantum system in a controlled fashion such that it displays properties not present in naturally occurring systems. Prominent examples are artificial lattices of atoms or molecules absorbed on surfaces\,\cite{gomes-12n306}, heterostructures made \eg from correlated-electron materials\,\cite{reyren-07s5842,chakhalian-14rmp1189}, and coupled-wire constructions of topological states of matter\,\cite{kane-02prl036401}.

A particularly interesting tool, applicable to bulk materials, is lattice strain which -- in a tight-binding description of electron dynamics -- induces inhomogeneous hopping energies. In the context of graphene, it has been theoretically shown\,\cite{PhysRevLett.101.226804,guinea-10np30,vozmediano-10pr109} that such inhomogeneous hopping mimics the effect of a vector potential in Dirac-fermion systems. If the resulting pseudo-magnetic field is sufficiently homogeneous -- applying \eg to triaxial strain patterns -- it can induce single-particle pseudo-Landau levels (PLLs) very similar to Landau levels in a physical magnetic field. Such PLLs have indeed been observed in strained graphene flakes\,\cite{levy-10s544} as well as in artificial molecular structures \cite{gomes-12n306}.  However, the resulting spectral quantization is approximate and restricted to energies near the Dirac point. Non-uniform strain has also been discussed for Weyl semimetals, but controlled effects are again restricted to the low-energy part of the spectrum\,\cite{weyl1,weyl2,weyl3}.

In this Letter, we lay out a scheme for strain engineering of single-particle levels which overcomes previous restrictions. We introduce tight-binding models with inhomogeneous hopping energies, defined on specific $d$-dimensional bipartite finite-size lattices, which display perfectly degenerate PLLs throughout their entire spectra. In $d=2$ our model resembles triaxial strain applied to the honeycomb lattice in the limit of strong electron-lattice coupling\,\cite{poli-14prb155418}, and we present the generalization to arbitrary $d$. Using iterative constructions, we are able to obtain the single-particle energies and their degeneracies in a closed algebraic form.
Most remarkably, our scheme paves the way to Landau level physics in three dimensions, realizable via tetraxial strain applied to the diamond lattice.


\paragraph{Model --}
Our tight-binding models are defined on a $d$-dimensional bipartite lattice, with sublattices A and B and coordination number $(d+1)$. The nearest-neighbor vectors $\dhat_j$ connect the center of a $(d+1)$-simplex to each of its vertices and, in the absence of strain, satisfy $\dhat_j^2=1$, $\dhat_j \cdot \dhat_{j'} = - \frac{1}{d}$ for $j\not= j'$, and $\sum_{j=1}^{d+1} \dhat_j=\bs 0$. 
Such $d$-dimensional bipartite lattices are referred to in the literature as hyperdiamond lattices\,\cite{ConwaySloane98,bedaque-08prd017502}.
In $d=1,2,3$ the relevant simplices are line segment, triangle, and tetrahedron, respectively.
The nearest-neighbor hopping Hamiltonian reads
\begin{equation}
\label{hh}
H=\sum_{\bs r \in{\rm B}}\sum_{j=1}^{d+1} t^{(N)}_{\bs r,j} \,c_{\bs r}^\dag\,c_{\bs r + \dhat_j} + {\rm H.c.}
\end{equation}
where $N=1,2,3,\ldots$ specifies the linear system size. In the presence of strain, we continue to use the coordinates of the unstrained lattice. Key ingredient are the inhomogeneous hopping amplitudes
\begin{equation}\label{ddim-strain}
t^{(N)}_{\bs r,j}={N-1-d\,{\bs r}\cdot{\dhat}_j\over d+1}\,;
\end{equation}
these can be generated by specific non-uniform strain in the limit of strong electron-lattice coupling\,\cite{supp}. In \eqref{ddim-strain} ${\bs r}\!=\!0$ defines the center of the system, and B sites are placed such that $t^{(N)}_{\bs r,j}$ is integer. The scalar product leads to a linear spatial variation of hopping amplitudes.
The hopping pattern \eqref{ddim-strain} is such that a set of $t^{(N)}_{\bs r,j}$ vanish identically, naturally cutting out a piece of size $N$, with the overall shape of the $d$-simplex, from a large lattice\,\cite{poli-14prb155418,supp}, see Fig.\,\ref{fig:1-2-3}.

As proven below, the level spectrum of $H$ reads
\begin{equation}\label{ddim-spec}
E_n^\pm = \pm\sqrt{N(N+d-2) - n(n+d-2)}
\end{equation}
for $n=1,\ldots, N-1$, and $E_N =0$ -- this is the central result of this Letter.
For $N\gg 1$ and $m\!=\!N\!-\!n\ll N$ the spectrum can be approximated by $\epsilon_m^\pm \equiv E_{N-m}^\pm \approx \pm\sqrt{2Nm}$. This corresponds to the low-energy spectrum of Dirac electrons subject to a vector potential, \ie the $\sqrt{m}$ behavior can be interpreted in terms of Dirac Landau levels in $d$ dimensions.
The degeneracy of each energy $E_n^\pm$ is given by
\begin{equation}\label{ddim-deg}
z_{d,n} = \frac{n(n+1)(n+2)\ldots (n+d-2)}{(d-1)!}\ .
\end{equation}
which is $\mathcal{O}(n^{d-1})$; specifically $z_{1,n}=1$, $z_{2,n}=n$, and $z_{3,n}=n(n+1)/2$.
In Fig.\,\ref{fig:1-2-3} we display the corresponding lattices as well as the level spectra for $N=4$.

We note that, to simplify expressions, Eq.\,\eqref{ddim-strain} is scaled to yield integer $t^{(N)}_{\bs r,j}$ and $E_n^2$; this results in a bandwidth $\propto N$. To obtain a spectrum with finite bandwidth for $N\to\infty$ requires to re-scale $t\to t/N$. Then, the low-energy levels follow $|\epsilon_m^\pm| \approx \sqrt{2m/N}$ corresponding to a pseudo-magnetic field which scales as $1/N$.

We now discuss the three important cases $d=1,2,3$ separately; in the remainder of this Letter we  prove the spectral properties for arbitrary $d$.

\paragraph{One-dimensional PLLs --}
In $d\!=\!1$ the discrete energy levels $E_n^\pm =\pm\sqrt{N(N-1) - n(n-1)}$ are non-degenerate, Eq.~\eqref{ddim-deg}. Nevertheless, the term ``Landau level'' appears justified given that the low-energy spectrum emulates a one-dimensional Dirac theory coupled to a vector potential of a homogeneous magnetic field.
The lattice, Fig.~\ref{fig:1-2-3}(a), is that of a two-atomic chain with an inhomogeneous Peierls-like distortion which increases away from the chain center. In analogy to $d\!=\!2,3$ one can interpret this hopping modulation as arising from biaxial strain, however, such a strain pattern cannot be realized using force fields in a solid. The $d\!=\!1$ case is mainly interesting as toy model.

\begin{figure}[b!]
\centering
\includegraphics[scale=0.35]{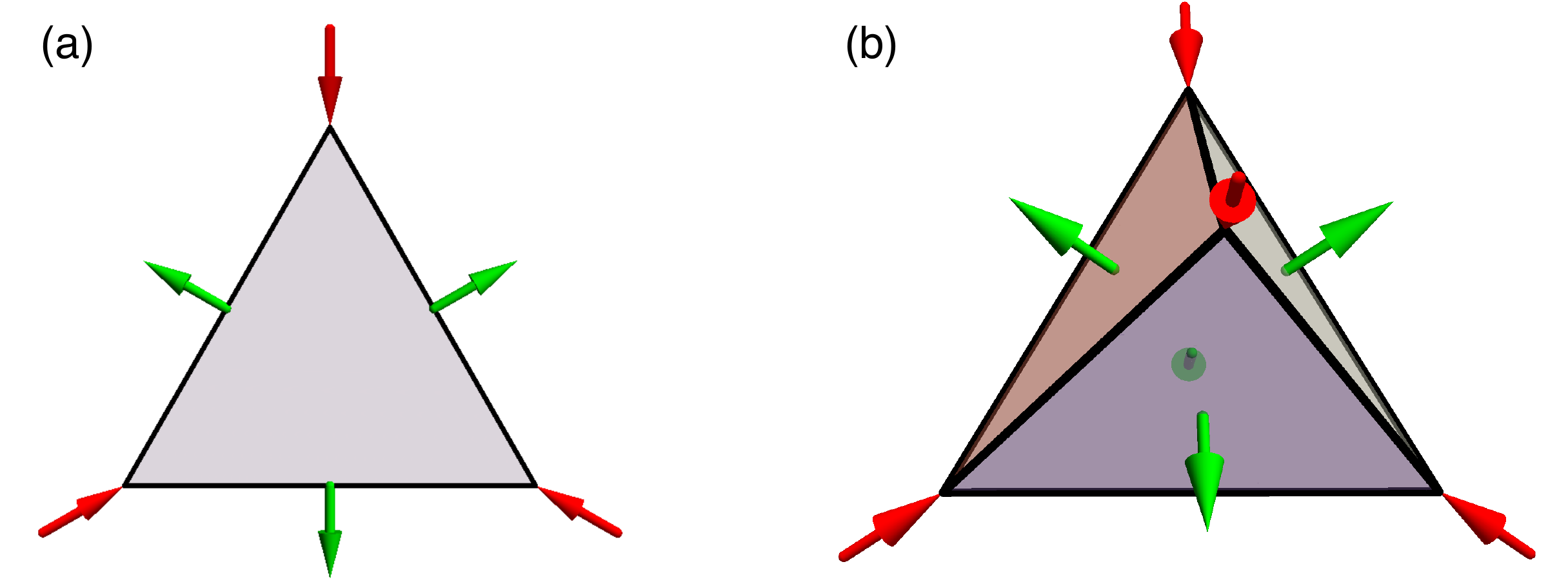} 
\caption{
Schematic illustration of forces for
(a) triaxial strain applied to a triangle and
(b) tetraxial strain applied to a tetrahedron.
The resulting displacement fields are (a) $\bs{u}^{\ }_{\rm 2D} = \bar C ( 2xy, x^2-y^2 )^T$ and (b)
$\bs{u}^{\ }_{\rm 3D} = \bar C ( yz, zx, xy )^T$ where $\bar C$ is a constant. 
}
\label{fig:tetraxial-strain}
\end{figure}

\paragraph{Two-dimensional PLLs --}
The $d\!=\!2$ case corresponds to triaxial strain, Fig.\,\ref{fig:tetraxial-strain}(a), applied to graphene\,\cite{guinea-10np30,vozmediano-10pr109,levy-10s544,gomes-12n306} in the limit of strong electron-lattice coupling\,\cite{poli-14prb155418,supp}. This limit, together with the specific sample shape, yields perfectly degenerate levels in the {\it entire} spectrum following the exact expression \eqref{ddim-spec}, $E_n^\pm = \pm\sqrt{N^2 - n^2}$, Fig.\,\ref{fig:1-2-3}\,(b). This is to be contrasted with PLLs for weak electron-lattice coupling which are smeared and restricted to low energies\,\cite{guinea-10np30,levy-10s544,gomes-12n306,neek-amal-13prb115428}. In the supplement\,\cite{supp} we illustrate the evolution between the two limits.

\paragraph{Three-dimensional PLLs --}
%
The most interesting case is $d\!=\!3$, which corresponds to {\it tetraxial} strain applied to a  diamond lattice. First, we recall that a tight-binding model (of $s$ orbitals, as opposed to the hybridized $sp^3$ orbitals of diamond) on the diamond lattice features a partial Dirac-type band touching at the $X$ points\,\cite{chadi-75pss405}. Second, the displacement vector for tetraxial strain, Fig.\,\ref{fig:tetraxial-strain}\,(b), is $\bs{u}^\pd_{\rm 3D} = \bar C ( yz, zx, xy)$. In the limit of strong electron-lattice coupling, tetraxial strain leads to hopping modulations as in Eq.\,\eqref{ddim-strain}.
The resulting spectrum $E_n^\pm = \pm\sqrt{N(N+1) - n(n+1)}$ is naturally interpreted as that of Landau levels in three spatial dimensions, see Fig.\,\ref{fig:1-2-3}\,(c). The corresponding continuum theory and its consequences will be published elsewhere\,\cite{elsewhere}. We have estimated that for realistic electron-lattice coupling 5\% strain should be sufficient to produce visible Landau levels\,\cite{supp}.
The construction of PLLs in $d\!=\!3$ is interesting on fundamental grounds, given that the quantum Hall effect exists only for even $d$; hence Landau levels in $d=3$ cannot be realized using physical magnetic fields.

%
%
\paragraph{Proof of the spectral properties --}
In order to prove Eqs.~\eqref{ddim-spec} and \eqref{ddim-deg}, we start with general properties of bipartite graphs and their implications. Then, we use these properties in combination with Eq.\,\eqref{ddim-strain} to iteratively construct the full spectrum. Further details of the proof are given in the supplement\,\cite{supp}.

\paragraph{Bipartite hopping and eigenmodes --}
Consider an arbitrary lattice, with all sites distributed into two sets called sublattices A and B. A hopping Hamiltonian whose only non-zero matrix elements connect A and B sites defines a bipartite hopping problem. The corresponding real Hamiltonian matrix $\mathcal{H}$ in a site basis, with A sites arranged before B sites, consists of off-diagonal blocks. As a result, the spectrum is particle-hole symmetric, \ie non-zero eigenvalues always come in pairs $E$ and $(-E)$.

If there is an imbalance of the number of sublattice sites, \ie $n_A \neq n_B$, there must be $|n_A- n_B|$ eigenvalues which vanish, $E=0$, with eigenvectors localized on sublattice $B$ if $n_B > n_A$\,\cite{zeronote}. The existence of zero modes was first noted by Sutherland\,\cite{sutherland86prb5208} and Lieb\,\cite{lieb89prl1201}; a simple proof is given in Ref.\,\onlinecite{inui-94prb3190}.

A further consequence of the bipartiteness is that the matrix $\mathcal{H}^2$ decouples into two disconnected blocks for the A and B sublattices. The eigenvalues of $\mathcal{H}^2$ are non-negative; the positive ones must be two-fold degenerate\,\cite{zeronote}, with one eigenvector solely defined on sublattice $A$ and the other one on sublattice $B$.

\paragraph{Notation ---}
We denote the Hamiltonian matrix of \eqref{hh} for linear size $N$ as $\mathcal{H}_N$. To establish  the spectrum \eqref{ddim-spec} we prove that $\mathcal{H}_N^2$ possesses eigenvalues $(E_n^\pm)^2 \equiv E_n^2$ being $2z_{d,n}$--fold  and $E_N^2 =0$  $z_{d,N}$--fold degenerate.
We denote the eigenvectors of $\mathcal{H}_N$ as $\psi^{(n,\mu)}_N$ and that of $\mathcal{H}_N^2$ as  $\phi^{(n,\nu)}_N$ where $\mu,\nu$ label degenerate eigenvectors. As noted, the positive-energy eigenvectors of $\mathcal{H}_N^2$ come in pairs, $\phi_{N,A}^{(n,\mu)}$ and $\phi_{N,B}^{(n,\mu)}$,  which have vanishing amplitude on sublattice $B$ and $A$, respectively. Bipartite hopping implies that $\mathcal{H}_N$ maps the two eigenstates onto each other\,\cite{signnote}:
\begin{equation}\label{AtoB}
\mathcal{H}^\pd_N\, \phi_{N,A}^{(n,\mu)} = E^\pd_n\,  \phi_{N,B}^{(n,\mu)} \,\,\hbox{and}\,\,
\mathcal{H}^\pd_N\, \phi_{N,B}^{(n,\mu)} = E^\pd_n\,  \phi_{N,A}^{(n,\mu)}\,.
\end{equation}
Note that \eqref{AtoB} is true for arbitrary bipartite hopping problems: It is a consequence of $\mathcal{H}_N$ connecting only $A$ and $B$ sites and the eigenvalue condition $\mathcal{H}_N^2 \,\phi^{(n)}_N = E_n^2 \,\phi^{(n)}_N$.

For further reference, we label the two non-zero blocks of the matrix $\mathcal{H}_N^2$ as $A_N$ and $B_N$, $\mathcal{H}_N^2=A_N \oplus B_N$, corresponding to its action on the A and B sublattices, respectively. Also, the sublattice with excess sites will be denoted B; the full expressions for the number of lattice sites for given $N$ and $d$ are provided in the supplement.


\paragraph{Iterative construction of the spectrum --}
{\em \underline{Step 1:}} For $N=1$ the lattice consists of a single B site, and we have $\mathcal{H}_1=\mathcal{H}_1^2=B_1=0$, $E_1=0$, and $\phi_1^{(N=1)}=\psi_1^{(N=1)}=1$.
(It will become clear below that $\phi_N^{(N)}=\psi_N^{(N)}$ for arbitrary $N$.)

{\em \underline{Step 2:}} We now show that the matrix $B_{N-1}$ is {\it identical} to $A_N$ up to a constant shift:
\begin{equation}\label{recursive}
A_N = B_{N-1} + \lambda(N)\cdot \mathbbm{1}
\end{equation}
where $\lambda(N) = 2N + d -3$ and $\mathbbm{1}$ is the corresponding unit matrix; the matrix dimensions of $A_N$ and $B_N$ are given in the supplement\,\cite{supp}.
As a result of Eq.~\eqref{recursive}, the eigenvectors of $B_{N-1}$ and $A_N$ are identical.

\begin{figure}[t!]
\centering
\includegraphics[scale=0.6]{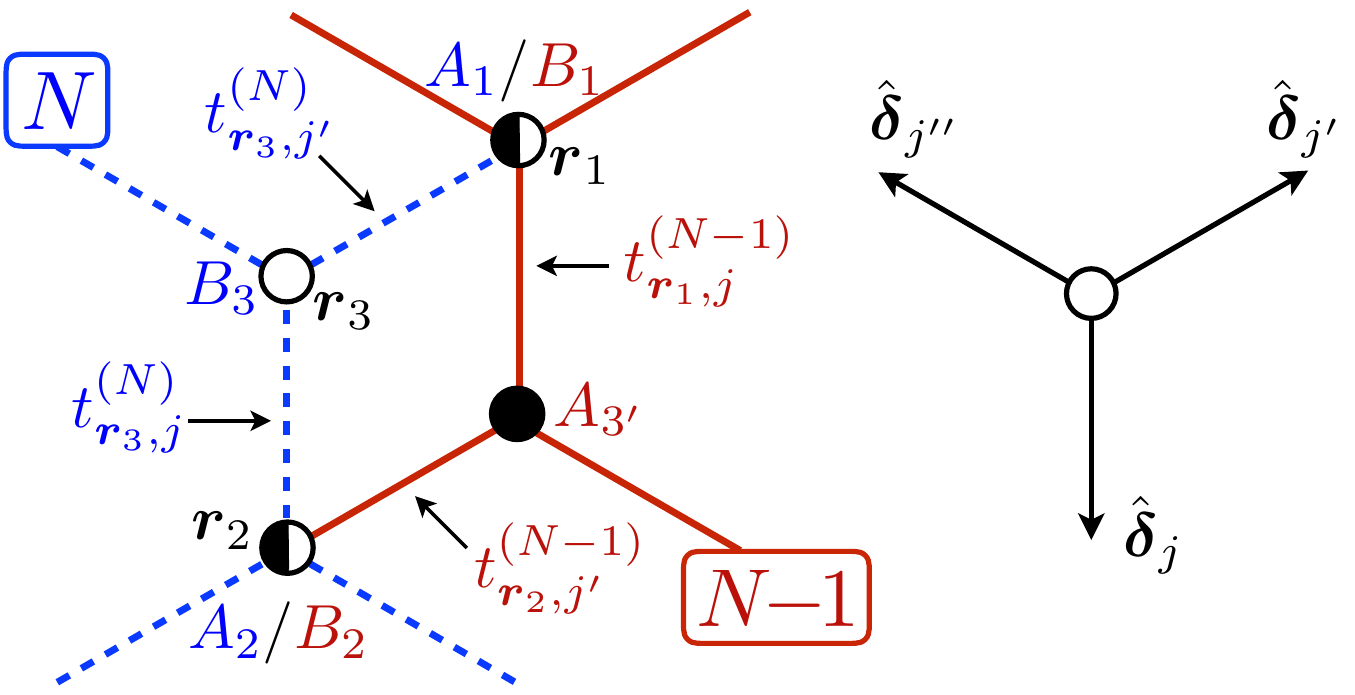}
\caption{
Illustration of the two different hopping paths (here exemplarily  $d=2$):
(i) the amplitude from $A_1$ to $A_2$ via $B_3$ along the blue bonds (for size $N$) is identical to (ii)
the amplitude from $B_1$ to $B_2$ via $A_{3'}$ along the red bonds (for size $N-1$). Filled (open) circles denote $A$ ($B$) sites.
}
\label{fig:1234}
\end{figure}

To prove Eq.~\eqref{recursive}, we first note that the positions of B sites for system size $N$ correspond to that of A sites for size $(N+1)$.
Second, B sites have neighboring sites in the $(d+1)$ $\dhat_j$ directions, while A sites have neighbors in the $(-\dhat_j)$ directions. Therefore the network of bonds between A and B sites has been locally inverted when switching from system size $N$ to $(N+1)$, see Fig.~\ref{fig:1234}.

Consider now a B site for size $N$ which is connected to its neighbors along the $(d+1)$ $\dhat_j$ directions via bonds of amplitudes $t^{(N)}_{\bs r,j} \in \mathbb{N}$. Then, the site with same coordinates for size $(N\!+\!1)$ (now belonging to sublattice A) is connected to its neighbors along the corresponding $(-\dhat_j)$ directions via bonds of amplitude $t^{(N+1)}_{{\bs r}',j}$ where ${\bs r}'={\bs r}-\dhat_j$. These hoppings obey $t^{(N+1)}_{{\bs r}',j} = t^{(N)}_{{\bs r},j} +1$ because
\begin{equation}
{N-d\,{\bs r}'\cdot\dhat_j\over d+1} -{N-1-d\,{\bs r}\cdot\dhat_j\over d+1} =1 \,.
\end{equation}

Next we determine the difference of the diagonal entries of the matrices $A_N$ and $B_{N-1}$. To that end, we abbreviate the $(d+1)$ hopping amplitudes surrounding an arbitrary A site for size $N$ as $t^{(N)}_1, t^{(N)}_2, \ldots,t^{(N)}_{d+1}$. They satisfy the sum rule $\sum_{j=1}^{d+1} t^{(N)}_j=N+d-1$\,\cite{supp}. Further, the site's diagonal entry into $A_N$ is given by $\sum_{j=1}^{d+1} (t^{(N)}_j)^2$. As noted, for size $(N\!-\!1)$ the hopping amplitudes surrounding the corresponding B satisfy $t^{(N-1)}_j = t^{(N)}_j-1$. Hence, the desired difference of the diagonal entries is
\begin{equation}
\sum_{j=1}^{d+1} \left[ (t^{(N)}_j )^2 - (t^{(N-1)}_j)^2\right] = \! \sum_{j=1}^{d+1} \left[ 2t^{(N)}_j - 1\right]= 2N +d-3\,.
\end{equation}

It remains to show that all off-diagonal entries in $A_N$ and $B_{N-1}$ are identical.
We consider two arbitrary adjacent B sites for size $(N\!-\!1)$ and label them as $B_1$ and $B_2$. For size $N$, the A sites with identical coordinates be $A_1$ and $A_2$; for an explicit illustration in $d=2$ see Fig.\,\ref{fig:1234}. The hopping from  $B_1$ to $B_2$ is via an A site ($A_{3'}$) along the trajectory $\dhat_j - \dhat_{j'}$, and the amplitude is $t^{(N-1)}_{\bs{r}_1,j}\, t^{(N-1)}_{\bs{r}_2,j'}$. As the network of $\dhat_j$ bonds between the sites has been locally inverted when switching from size $(N\!-\!1)$ to $N$, the hopping from $A_1$ to $A_2$ is via a B site ($B_3$) along the trajectory $- \dhat_{j'} + \dhat_j$, with amplitude  $t^{(N)}_{\bs{r}_3,j'}\, t^{(N)}_{\bs{r}_3,j}$.
From Fig.\,\ref{fig:1234} we see that $\bs{r}_1=\bs{r}_3+\dhat_{j'}$ and $\bs{r}_2=\bs{r}_3+\dhat_{j}$. Using  $\dhat_j\cdot\dhat_{j'} = -\frac{1}{d}$ and Eq.\,\eqref{fig:1234} we obtain $t^{(N-1)}_{\bs{r}_1,j} = t^{(N)}_{\bs{r}_3,j}$ and $t^{(N-1)}_{\bs{r}_2,j'} = t^{(N)}_{\bs{r}_3,j'}$. Since the considered hoppings are identical, the pairwise products entering the matrices $A_N$ and $B_{N-1}$ are identical, too. This concludes the proof of Eq.\,\eqref{recursive}.

{\em \underline{Step 3:}}
A direct consequence of Eq.~\eqref{AtoB} and the Lieb-Sutherland theorem is that the spectrum of $B_N$ is given by the one of $A_N$ plus $z_{d,N}$ zero eigenvalues, where $z_{d,N}$ is the number of excess B sites. Specifically, we have $z_{1,N}=1$, $z_{2,N}=N$, and $z_{3,N}=N(N+1)/2$; the general expression for $z_{d,N}$ \eqref{ddim-deg} will be derived in the supplement\,\cite{supp}.
Note that  the eigenvector structure of the null space (\ie the zero modes) is non-trivial and generally not known.

{\em \underline{Step 4:}} From Eq.\,\eqref{recursive} it follows that
\begin{equation}
E_n^2(N) = \sum_{\mu=n+1}^N \lambda(\mu) = N(N+d-2) - n(n+d-2)
\end{equation}
and we obtain the previously proposed spectrum of $\mathcal{H}_N^2$.

{\em \underline{Step 5:}}  Finally we show that the degeneracies of the levels at non-zero energy are indeed given by \eqref{ddim-deg}. This follows from the iterative construction: The $z_{d,N}$ zero modes of size $N$ become, upon increasing the size to $(N\!+\!1)$, the $2z_{d,N}$ states with energy $E_N^2(N+1)=\lambda(N+1)$ ($z_{d,N}$ states on each of the sublattices A and B), while there are $z_{d,N+1}$ zero modes etc.
As a cross-check we determine the total number of states for size $N$: This is  $z_{d,N}+\sum_{n=1}^{N-1}2z_{d,n}$ which equals the number of lattice sites, $M_{d,N}$, and thus the Hilbert space dimension\,\cite{supp}. Hence, there cannot be additional energy levels. This completes the proof.

\begin{figure}[t!]
\centering
\includegraphics[scale=0.46]{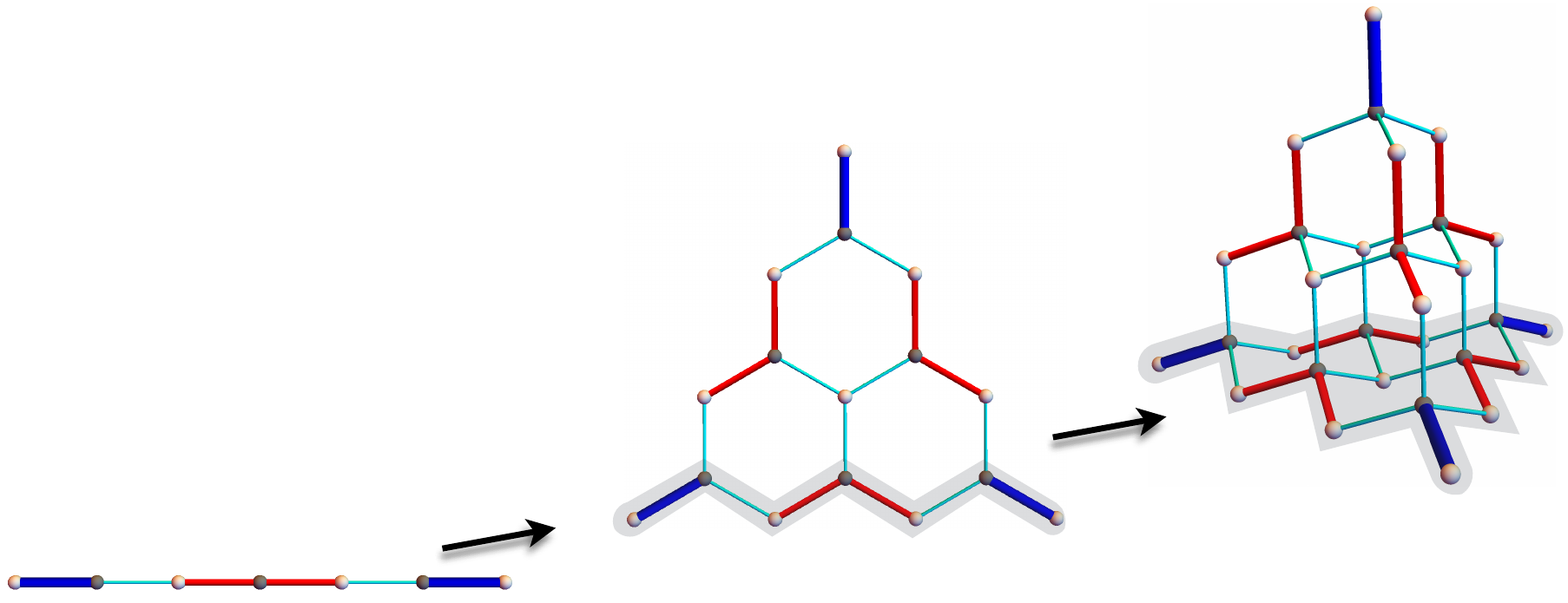}
\caption{Dimensional iteration of the $d$-dimensional simplices: the $d$-dimensional simplex is constructed from $N$ $(d\!-\!1)$-dimensional simplices. For instance, the triangle ($d\!=\!2$) consists of a chain ($d\!=\!1$) with $N=4$, one with $N=3$, one with $N\!=\!2$, and one with $N\!=\!1$. Correspondingly, the tetrahedron ($d\!=\!3$) consists of several triangles. This dimensional dependence manifests itself in the expressions for  degeneracies $\eqref{ddim-deg}$ and number of lattice sites\,\cite{supp}; see text.}
\label{fig:dim-iteration}
\end{figure}

%
%

\paragraph{Dimensional hierarchy --}
It is instructive to consider a dimensional iteration, Fig.~\ref{fig:dim-iteration}.
A chain ($d\!=\!1$) of size $N$ has one zero mode and consists of $(2N\!-\!1)$ sites.
A triangle ($d\!=\!2$) of size $N$ consists of $N$ chains (all with different length, but one excess B site), thus there are $N$ zero modes and $\sum_{j=1}^N (2j\!-\!1) = N^2$ sites.
Similarly, a tetrahedron ($d\!=\!3$) of size $N$ consists of $N$ triangles (all with different size), yielding $\sum_{j=1}^N j = N(N+1)/2$ zero modes and $\sum_{j=1}^N j^2 = N(N+1)(2N+1)/6$ sites.
This dimensional iteration allows us to recover the general results for $z_{d,N}$ \eqref{ddim-deg} and for $M_{d,N}$, as detailed in the supplement\,\cite{supp}.
It also enables a hierarchical construction of the lattices, i.e., of the vectors $\dhat_j$.
In $d=1$, the two nearest-neighbor displacements are $\dhat_{1,2}=\pm\hat{\bs x}$.
In $(d\!-\!1)$ dimensions, the $d$ unit vectors $\dhat_j$ satisfy $\dhat_j\cdot\dhat_{j'}=-1/(d-1)$ for $j\ne j'$.  Now define $\dhat'_j=\sqrt{1-d^{-2}}\,\dhat_j - d^{-1}\,\hat{\bs e}^\pd_d$ and $\dhat'_{d+1}=\hat{\bs e}^\pd_d$, where $\hat{\bs e}^\pd_d$ is the unit vector in the $d^{\rm th}$ dimension.  Then $\big\{\dhat'_j,\ldots,\dhat'_{d+1}\big\}$ are the nearest-neighbor unit vectors in $d$ dimensions.

\paragraph{Conclusion --}
We have introduced and analyzed a family of tight-binding models for specific finite-size lattices in arbitrary dimension $d$ where strain-induced inhomogeneous hopping leads to perfectly degenerate PLLs. This is remarkable in two respects:
(i) Degenerate Landau levels usually only follow from (approximate) continuum theories, while lattice models yield only approximately degenerate Landau levels. In contrast, here we have a lattice realization of perfect degeneracies.
(ii) While Landau levels in $d\!=\!2$ may be realized using either magnetic field or strain, there is no magnetic-field route in $d\!=\!3$. Hence, our PLL construction provides a unique way of obtaining perfectly flat bands in three dimensions.
Upon including electron-electron interactions, flat bands open the exciting possibility to study fractionalization in three spatial dimensions, similar to what has been done in $d\!=\!2$ for strained graphene\,\cite{ghaemi-12prl266801,abanin-12prl066802,venderbos-16prb195126}. This will be the subject of future work. We note that pseudo-magnetic fields and the associated pseudo-Landau levels in $d=1, 2, 3$ could in principle be realized in cold-atom settings\,\cite{bakr-09n74,muldoon-12njp073051} which also enable  tunable interactions.


\acknowledgements

We thank C.\ Poli, H.\ Schomerus, W.\ Lang, and C.\ Timm for discussions and L.\ Fritz for previous collaborations on related topics.
%
This research was supported by the DFG through SFB 1143, SPP 1666, and GRK 1621
as well as by the Helmholtz association through VI-521. 
DPA is grateful to the hospitality of the ITP and SFB 1143 at TU Dresden.


\bibliographystyle{prsty}
\bibliography{triangles}

\begin{thebibliography}{10}
\bibitem{zeronote}
If the number of zero modes exceeds $|n_B - n_A|$ then the additional number of
  zero modes must be even.
\bibitem{signnote}
The normalized real eigenvectors $\phi_{A}$ and $\phi_{B}$ are defined up to an
  overall sign. A change $\phi_{A}\to-\phi_{A}$ or $\phi_{B}\to-\phi_{B}$ leads
  to $E_n\to-E_n$ in Eq.~\eqref{AtoB}.
\end{thebibliography}

\end{document}


\title{Supplemental material for\\
Strain-induced Landau Levels in arbitrary dimensions with an exact spectrum}

\author{Stephan Rachel}
\affiliation{Institut f\"ur Theoretische Physik, Technische Universit\"at Dresden,
01062 Dresden, Germany}
\author{Ilja G\"othel}
\affiliation{Institut f\"ur Theoretische Physik, Technische Universit\"at Dresden,
01062 Dresden, Germany}
\author{Daniel P.\ Arovas}
\affiliation{Department of Physics, University of California, San Diego, La Jolla, California 92093, USA}
\author{Matthias Vojta}
\affiliation{Institut f\"ur Theoretische Physik, Technische Universit\"at Dresden,
01062 Dresden, Germany}


\maketitle

\section{Shape of finite-size lattices}

\begin{figure}[h!]
\centering
\includegraphics[scale=0.9]{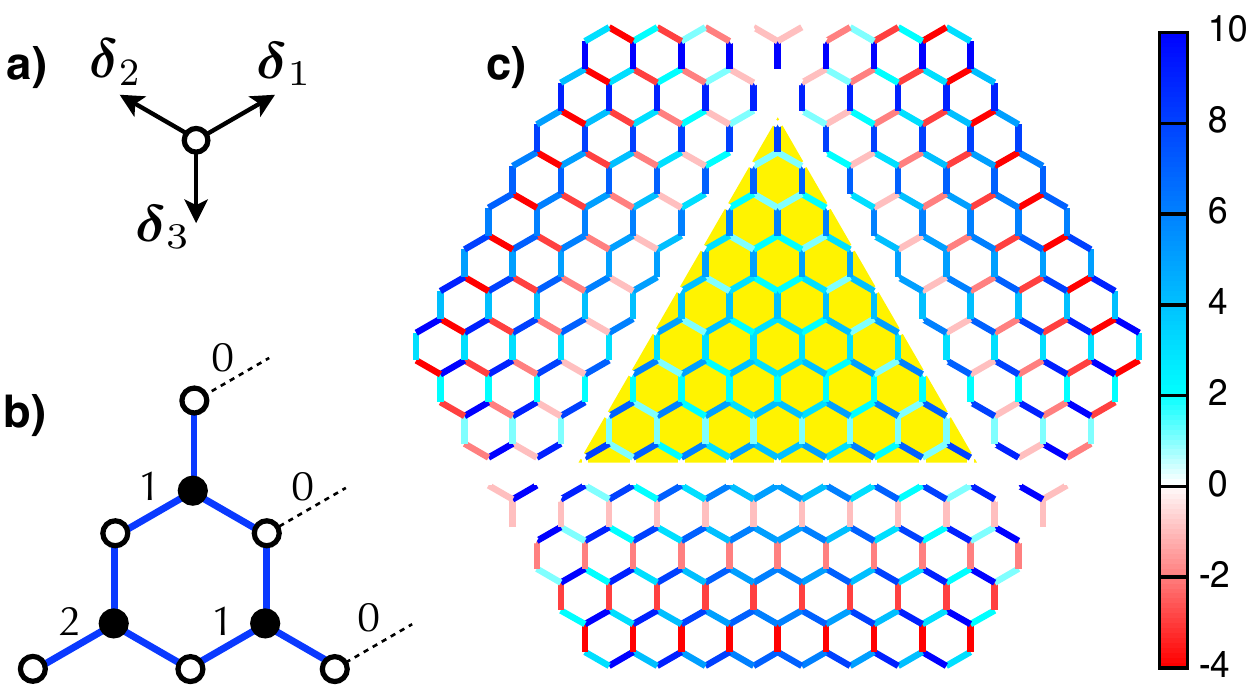}
\caption{(a) Definitions of the nearest-neighbor vectors $\bs{\delta}_j$ $(j=1,2,3)$. (b) A honeycomb-lattice triangle of size $N=3$ with the hopping energies $t^{(N=3)}_{j=1}$ along the $\bs{\delta}_1$ direction. (c) Due to the linear hopping modulation {\eqhopping} a triangle (highlighted in yellow) is naturally cut out off a larger honeycomb lattice. The bond colors correspond to the hopping value $t^{(N=9)}_{\bs{r},j}$. A figure similar to panel (c) appears in Ref.~\onlinecite{poli-14prb155418}.
}
\label{fig:figure1}
\end{figure}

In this section, we explicate the relation between the hopping amplitudes, Eq.\,{\eqhopping} in the main paper, and the particular shape of the considered finite-size lattices in $d$ dimensions. The scalar product in Eq.\,{\eqhopping} leads to a purely ``linear'' hopping modulation, \ie along a certain $\dhat_j$ direction, bonds which are parallel to  $\dhat_j$ and which possess the same component along the $\dhat_j$ direction must have the same hopping amplitude. As an example, in Fig.\,\ref{fig:figure1}\,(b) a triangle of size $N=3$ is shown. Bonds along the $\dhat_1$ direction are then associated with a hopping amplitude 2, the next ``row of bonds'' with 1, and the next ``row of bonds'' would be associated with a hopping amplitude 0. Eq.\,{\eqhopping} is chosen such that these zero bonds occur in all $\dhat_j$ directions, such that a finite-size system is naturally cut out off a larger lattice.\cite{poli-14prb155418}

This is illustrated in Fig.\,\ref{fig:figure1}\,(c) for a honeycomb lattice ($d=2$). Implementing the hopping modulation Eq.\,{\eqhopping}, with $N=9$ and ${\bs r}=0$ placed in the center of a hexagon, results in several disconnected pieces of a large (or infinite) lattice. We choose the yellow triangle in the center as the finite-size lattice of interest. In $d=1$ ($d=3$) a chain (tetrahedron) is cut out off a larger embedding lattice in the same way as here described for the honeycomb lattice.

\section{Number of  sites and zero modes through dimensional hierarchy}\label{sec:numberofsites}

Once the size $N$ and shape of a finite-size lattice are determined, the total number of lattice sites can be deducted based on geometrical reasoning. Also the degeneracies as given in Eq.\,{\eqdeg} can be geometrically deducted for a given $d$ because each energy subspace corresponds for smaller size $N$
to the null space containing the zero modes. As pointed out in the main text, the zero modes are a consequence of the imbalance of $A$ and $B$ sites (Lieb-Sutherland theorem). This imbalance of $A$ and $B$ sites is again a consequence of geometry.

The simplices considered in this paper are related to each other by a dimensional hierarchy, as pointed out in the main text. We now employ this hierarchy to iteratively obtain both the total number of lattice sites as well as the degeneracies of zero modes.

\subsection{Number of lattice sites}

For $d=1$, the smallest system size $N=1$ consists of a single $B$ site. Increasing the size $N$ by one results in two additional lattice sites, an $A$ and a $B$ site. That is, the number of lattice sites for size $N$ is given by
\begin{equation}
M_{1,N} = 2N-1\ .
\end{equation}
The $d=2$ simplex (triangle) of size $N$ consists of $N$ $d=1$ simplices (chains), each with different size $\mu$: a $d=1$ simplex with size $\mu=1$, a $d=1$ simplex with size $\mu=2$, \ldots, and finally a $d=1$ simplex with size $\mu=N$ [see Fig.\,{\figiteration}]. Thus we can write for the total number of sites in $d=2$
\begin{equation}
M_{2,N} = \sum_{\mu=1}^N 2\mu-1 = N^2\ .
\end{equation}
The $d=3$ simplex (tetrahedron) of size $N$ consists of $N$ $d=2$ simplices, each with different size $\mu$: a $d=2$ simplex with size $\mu=1$, a $d=1$ simplex with size $\mu=2$ (\ie $\mu^2=4$ sites), \ldots, and finally a $d=1$ simplex with size $\mu=N$ (\ie $\mu^2=N^2$ sites). Thus we can write for the total number of sites in $d=3$
\begin{equation}
M_{3,N} = \sum_{\mu=1}^N \mu^2 = \frac{N(N+1)(2N+1)}{6}\ .
\end{equation}
Continuing this procedure results in
\begin{equation}\begin{split}
&M_{4,N} = \frac{N(N+1)^2(N+2)}{12}\ , \qquad M_{5,N}=\frac{N(N+1)(N+2)(N+3)(2N+3)}{120}\ , \\[5pt]
&M_{6,N} = \frac{N(N+1)(N+2)^2(N+3)(N+4)}{360}\ ,\qquad M_{7,N}=\frac{N(N+1)(N+2)(N+3)(N+4)(N+5)(2N+5)}{5040}\ .
\end{split}
\end{equation}
By induction one can easily find the result for general $d$:
\begin{equation}
\label{ddim-sites}
M_{d,N} = 
 \frac{(2N+d-2)N(N+1)(N+2)\ldots (N+d-2)}{d!}\ ,  \\[5pt]
\end{equation}

\subsection{Number of zero modes}

In order to obtain the degeneracies $z_{d,N}$, Eq.~{\eqdeg} in the main paper, from this dimensional hierarchy, we recall that each $d=1$ simplex contributes one zero mode. Since the $d=2$ simplex consists of $N$ chains, we find the number of zero modes as
\begin{equation}
z_{2,N}=\sum_{\zeta=1}^N 1 = N\ .
\end{equation}
As suggested by Fig.\,{\figiteration} and mentioned before, the $d=3$ simplex consists of $N$ $d=2$ simplices each of which contributes $\zeta$ zero modes ($\zeta=1,\ldots, N$):
\begin{equation}
z_{3,N}=\sum_{\zeta=1}^N \zeta = \frac{N(N+1)}{2}\ .
\end{equation}
Continuing this procedure yields
\begin{equation}
z_{4,N}=\sum_{\zeta=1}^N \frac{\zeta(\zeta+1)}{2} = \frac{N(N+1)(N+2)}{6}\ ,\qquad
z_{5,N}=\sum_{\zeta=1}^N \frac{\zeta(\zeta+1)(\zeta+2)}{6} = \frac{N(N+1)(N+2)(N+3)}{24}\,.
\end{equation}
The generalization to arbitrary dimension $d$ is obvious and results in Eq.\,{\eqdeg}.


\section{Matrix dimensions}\label{sec:matrixdimension}

Both $\mathcal{H}_N$ and $\mathcal{H}_N^2$ are matrices of size $M_{d,N}\times M_{d,N}$, see Sec.\,\ref{sec:numberofsites}. As mentioned previously, $\mathcal{H}_N^2$ is a block-diagonal matrix consisting of two disconnected blocks, $\mathcal{H}_N^2=A_N \oplus B_N$.
$B_N$ has the same spectral content as $A_N$ except for the $z_{d,N}$ zero modes. Consequently, the matrix dimensions are
\begin{equation}
{\rm dim}(A_N) = \frac{1}{2}\left( M_{d,N} - z_{d,N} \right) \qquad \hbox{and} \qquad
{\rm dim}(A_N) = \frac{1}{2}\left( M_{d,N} + z_{d,N} \right)
\end{equation}
Explicitly for the most relevant cases $d=1,2,3$ we obtain
\begin{align}
d=1:\quad  & {\rm dim}(A_N) = N - 1\,, \qquad\qquad\qquad  {\rm dim}(B_N) = N\ , \\[10pt]
d=2:\quad  & {\rm dim}(A_N) = \frac{ (N-1)N}{2}\,, \qquad\qquad  {\rm dim}(B_N) = \frac{N(N+1)}{2}\ , \\[10pt]
d=3:\quad  & {\rm dim}(A_N) = \frac{(N-1)N(N+1)}{6}\,, ~~ {\rm dim}(B_N) = \frac{N(N+1)(N+2)}{6}\ .
\end{align}

\vspace{10pt}


\section{Sum rules}

In the main text, it has been stated that the hopping amplitudes $t^{(n)}_{j}$ associated with the $(d+1)$ bonds surrounding an $A$ lattice site of a $d$ simplex fulfill the following sum rule,
\begin{equation}\label{sum-rule-A}
{^{[A]}}\sum_{j=1}^{d+1} t_j^{(N)} = N + d -1\ .
\end{equation}

\begin{figure}[h!]
\centering
\includegraphics[scale=0.67]{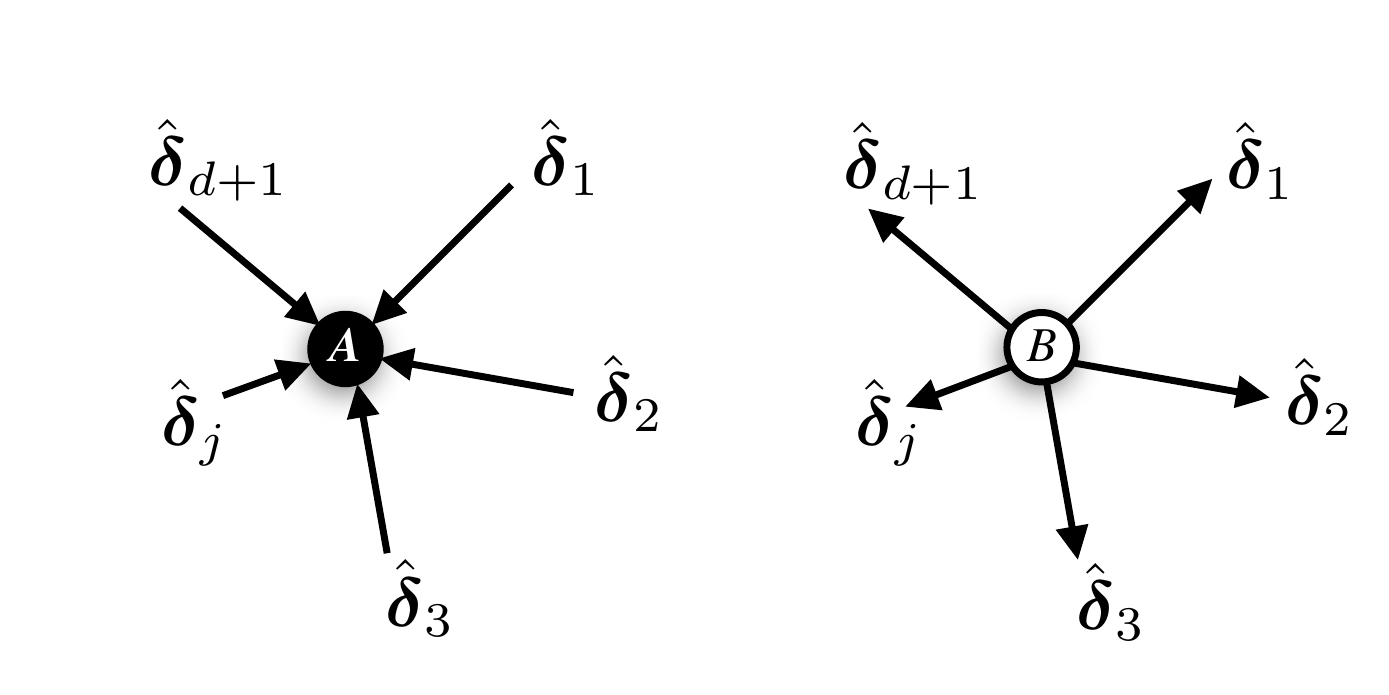}
\caption{(Left) $A$ site of a $d$ simplex with its $(d+1)$ nearest-neighbor vectors. (Right) The analogous $B$ site of a $d$ simplex has the same $(d+1)$ nearest neighbor vectors but with opposite orientation.}
\label{fig:A-B_vs_delta}
\end{figure}

In order to prove Eq.\,\eqref{sum-rule-A}, it is instructive to consider Fig.\,\ref{fig:A-B_vs_delta}. Let us introduce the position vector of the considered $A$ site as $\bs{r}_0$. Recalling that
the definition of the hopping amplitudes, Eq.\,{\eqhopping} in the main text, refers to ${\bs r \in{\rm B}}$, we note that the $(d+1)$ B-sublattice neighbors of $\bs{r}_0$ are $\bs{r}_j = \bs{r}_0 -\dhat_j$ ($j=1,\ldots, d+1$). Then we can write the sum as follows:
\begin{eqnarray}
{^{[A]}}\sum_{j=1}^{d+1} t_j^{(N)} &=& \sum_{j=1}^{d+1} t^{(N)}_{\bs{r}_j,j}  =  \sum_{j=1}^{d+1} \frac{N-1-d \,\bs{r}_j\cdot\dhat_j}{d+1}\\[5pt]
&=& N-1 - \frac{d}{d+1} \sum_{j=1}^{d+1} \bs{r}_j \cdot \dhat_j \\[5pt]
&=& N-1 -\frac{d}{d+1} \sum_{j=1}^{d+1} \Big\{ \bs{r}_0 \cdot \dhat_1 + \bs{r}_0\cdot\dhat_2 + \ldots + \bs{r}_0\cdot\dhat_{d+1} - d - 1 \Big\} \\[5pt]
&=& N-1 + d \quad\boxed{}
\end{eqnarray}
Here we used the vector relation $\sum_{j=1}^{d+1} \dhat_j =\bs{0}$ of a $d$-simplex.

\vspace{10pt}
While not used in the main paper, the analogous sum rule for $B$ sites reads
\begin{equation}\label{sum-rule-B}
{^{[B]}}\sum_{j=1}^{d+1} t_j^{(N)} = N -1\ ,
\end{equation}
which is independent of $d$. The proof is even simpler than before (and there is no need to introduce other positions than $\bs{r}_0$):
\begin{eqnarray}
{^{[B]}}\sum_{j=1}^{d+1} t_j^{(N)} &=& \sum_{j=1}^{d+1} t^{(N)}_{\bs{r}_0,j}  = N-1-\frac{d}{d+1}\sum_{j=1}^{d+1} \bs{r}_0\cdot \dhat_j  = N-1 \quad\boxed{}
\end{eqnarray}

\section{Iterative construction of eigenstates of $\bs{\mathcal{H}_N^2}$}

The central observation of the iterative spectral procedure for arbitrary $d$, discussed previously and in the main paper, is based on the fact that each energy subspace $E_n^2(N)$ [throughout this section we add the label $N$ in parentheses to $E_n^2$ explicitly] originates from the null space $E_{n}^2(n)$, \ie the zero modes, of a smaller system with size $n$.

\begin{figure}[t!]
\centering
\includegraphics[scale=1.3]{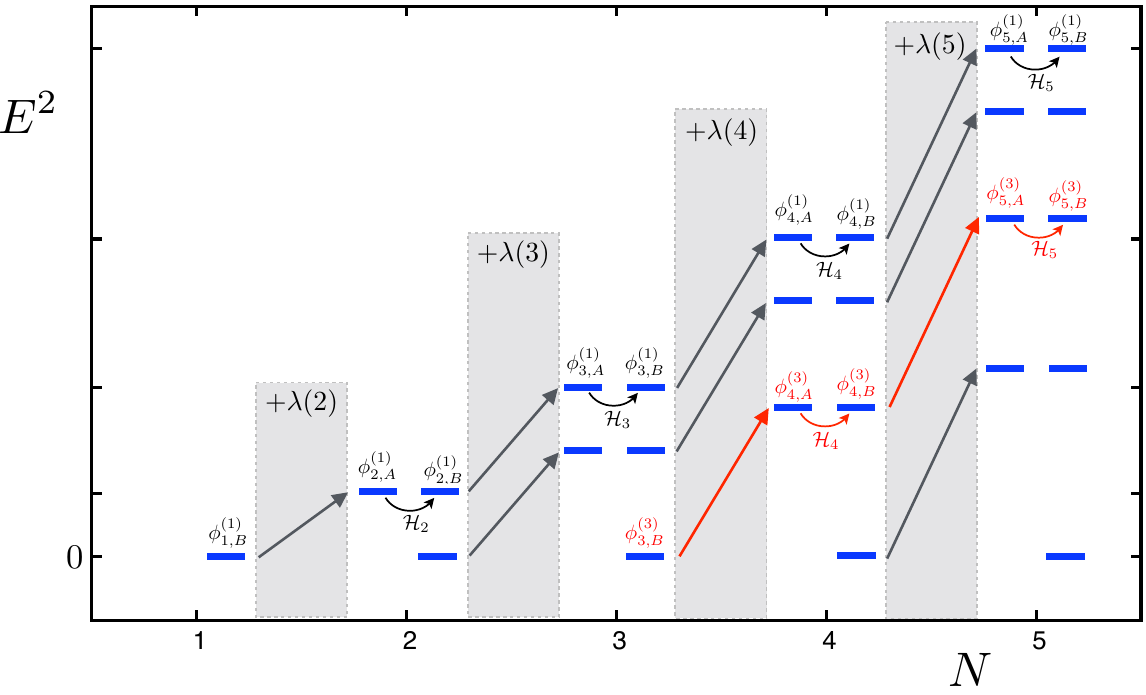}
\caption{
Iterative structure of spectrum: each energy subspace at system size $N$ corresponds to a null space of a smaller system. Degeneracy labels of the $\phi_{N}^{(n)}$ are suppressed for the sake of clarity.
The spectral shift according to Eq.~\eqref{recursive} is visualized as the grey boxes (``$+\lambda(N)$''), and the switching from $A_N$ to $B_N$ according to Eq.~\eqref{AtoB} is indicated explicitly by the round arrows with the corresponding label $\mathcal{H}_N$.
The red labels and arrows refer to the example in Eq.~\eqref{example}.
}
\label{fig:iterative}
\end{figure}

This can be best seen from Eqs.\,{\eqHAB} and {\eqfive} of the main paper, which we reproduce below for reference. The matrix relation
\begin{equation}\label{recursive}
A_N = B_{N-1} + \lambda(N)\cdot \mathbbm{1}
\end{equation}
implies that the eigenstates of $A_N$ (which are also the eigenstates $\phi^{(n)}_{A,N}$ of $\mathcal{H}_N^2$ associated with sublattice $A$, embedded into the larger Hilbert space) are the same as the one of $B_{N-1}$, since a constant shift does not affect the eigenstates, only the spectrum.
In 
\begin{equation}\label{AtoB}
\mathcal{H}^\pd_N\, \phi_{N,A}^{(n,\mu)} = E^\pd_n\,  \phi_{N,B}^{(n,\mu)} \,\,\hbox{and}\,\,
\mathcal{H}^\pd_N\, \phi_{N,B}^{(n,\mu)} = E^\pd_n\,  \phi_{N,A}^{(n,\mu)}\,.
\end{equation}
we have then the opposite situation: when applying $\mathcal{H}_N$ to $\phi^{(n)}_{A,N}$ the Hilbert space is effectively increased (namely resulting in $B_N$) and the eigenvectors are changed, but the spectrum remains unchanged.
We conclude that all eigenvectors of $\mathcal{H}_N^2$ are effectively generated by applying $\mathcal{H}_N$ several times (but each time for different $N$) within this iterative construction. These considerations are illustrated in Fig.\,\ref{fig:iterative}.

We will now use this insight to explicitly express the eigenstates $\phi_N^{(n)}$ through the zero modes $\phi_{n,B}^{(n)}$. First, we recall that in the step where $\mathcal{H}_N$ is applied to $\phi^{(n)}_{A,N}$, the effective Hilbert space is increased (namely from ${\rm dim}(A_N)$ to ${\rm dim}(B_N)$). Therefore we cannot directly write \eg $\phi_{5,B}^{(1)} \propto \mathcal{H}_5 \mathcal{H}_4 \mathcal{H}_3 \mathcal{H}_2 \phi_{1,B}^{(1)}$ because the  $\mathcal{H}$ matrices have different dimension. However, we can rephrase the multiple application of $\mathcal{H}$ using its block structure:
The bipartiteness of the hopping implies
\begin{equation}
\mathcal{H}_N = \left( \begin{array}{cc}~\bs{0} ~~~~~\mathcal{M}_N~~ \\[5pt]
\mathcal{M}_N^T ~~~\bs{0}~~\\[2pt]
\end{array}\right)
\end{equation}
where $\mathcal{M}_N$ is a rectangular ${\rm dim}(A_N)\times{\rm dim}(B_N)$ matrix, see Sec.\,\ref{sec:matrixdimension}.
Apparently the $\mathcal{M}_N^T$ are the relevant blocks of $\mathcal{H}_N$  used in the iterative construction when switching from $A_N$ to $B_N$.
We previously introduced eigenvectors $\phi_{N,A/B}^{(n)}$ of $\mathcal{H}_N^2$ which are non-zero only on either sublattice A or B, respectively. (We suppress the degeneracy label of the eigenvector here and in the following, as this appears identically of both sides of the equations.)
Now we also introduce eigenvectors of $A_N$ and $B_N$:
\begin{equation}
A_N {\tilde\phi}_{A,N}^{(n)} = E_n^2(N) {\tilde\phi}_{A,N}^{(n)} \qquad \hbox{and} \qquad
B_N {\tilde\phi}_{B,N}^{(n)} = E_n^2(N) {\tilde\phi}_{B,N}^{(n)}\ .
\end{equation}
Note that the $\tilde\phi$ are isomorphic to the $\phi$ vectors,
\begin{equation}
\phi_{N,A} = \left(\begin{array}{c}\\[2pt]{\tilde\phi}_{N,A}\\[13pt]
0\\[-3pt] \vdots \\[-3pt] \vdots \\0 \\  \end{array}\right) \qquad \hbox{and} \qquad
\phi_{N,B} = \left(\begin{array}{c} 0 \\[-5pt] \vdots \\ 0 \\[20pt] {\tilde\phi}_{N,B}\\[20pt] \end{array}\right)
\end{equation}
Now we are prepared to write the previous example correctly as ${\tilde\phi}_{5,B}^{(1)} \propto \mathcal{M}_5^T \mathcal{M}_4^T \mathcal{M}_3^T \mathcal{M}_2^T {\tilde\phi}_{1,B}^{(1)}$.

The explicit construction of eigenstates of $\mathcal{H}_N^2$ is thus based on the equivalence of eigenstates ${\tilde\phi}_{N-1,B}^{(n)} \equiv {\tilde\phi}_{N,A}^{(n)}$ according to Eq.~\eqref{recursive}, and the $A\to B$ switch $\mathcal{M}_N^T {\tilde\phi}_{A,N}^{(n)} = E_n(N) \,{\tilde\phi}_{B,N}^{(n)}$ which follows from Eq.~\eqref{AtoB}.
Then, arbitrary eigenstates ${\tilde\phi}_{N,B}^{(n)}$ can be written as
\begin{equation}\label{general_phi_BNn}
{\tilde\phi}_{N,B}^{(n)} = \frac{1}{\gamma_n^{(n+1,N)}} \Big( \prod_{j=n+1}^N \mathcal{M}_j \Big)^T \,\, {\tilde\phi}_{n,B}^{(n)} \,
\end{equation}
while eigenstates ${\tilde\phi}_{N,A}^{(n)}$ are given by
\begin{equation}\label{general_phi_ANn}
{\tilde\phi}_{N,A}^{(n)} = \frac{1}{\gamma_n^{(n+1,N-1)}} \Big( \prod_{j=n+1}^{N-1} \mathcal{M}_j \Big)^T \,\, {\tilde\phi}_{n,B}^{(n)}\ .
\end{equation}
where the normalization factor is
\begin{equation}\label{gamma}
\gamma_n^{(\mu,\nu)} = \prod_{j=\mu}^\nu E_n (j)\ .
\end{equation}

We illustrate the previous findings by giving a simple example: the eigenstate $\phi_{5,B}^{(3)}$ corresponding to the energy $E_3^2(5)$ originates from the null space for size $N=3$, $\phi_{3,B}^{(3)}$, and can thus be expressed as
\begin{equation}
\label{example}
{\tilde\phi}_{5,B}^{(3)} =  \frac{1}{\gamma_3^{(4,5)}} M_5^T \cdot M_4^T \,\, {\tilde\phi}_{3,B}^{(3)}
\end{equation}
with $\gamma_3^{(4,5)} = E_3(4) \cdot E_3(5) = \sqrt{2(5+d)(6+d)}$.
This example is illustrated by the red labels and arrows in Fig.\,\ref{fig:iterative}.


\section{Strain and modulated hoppings}

The model presented in the paper employs an explicit modulation of the hopping energies, without reference to actual lattice distortions. In this section, we will discuss the relation between strain and hopping modulation in more detail and, in particular, discuss the limit of strong electron-lattice coupling alluded to in the paper.

\subsection{Electron-lattice coupling and ``linear'' hopping modulation}\label{sec:VIA}

Consider a regular atomic lattice subject with a strain-induced distortion described by the displacement field $\bs{u}$ which is related to the strain tensor $\overline{\overline{U}}$ via $\overline{\overline{U}} = [\bs{\nabla} \bs{u} + (\bs{\nabla} \bs{u})^T]/2$.
For the concrete cases of triaxial and tetraxial strain relevant to the paper, we have $\bs{u}^{\ }_{\rm 2D} = \bar C/a_0\, ( 2xy, x^2-y^2 )^T$ and $\bs{u}^{\ }_{\rm 3D} = \bar C/a_0\, ( yz, zx, xy )^T$, respectively, where $\bar C$ quantifies the distortion of the lattice, and $a_0$ is the original lattice constant which we introduce here explicitly (in the rest of the supplement and in the main paper we set $a_0 \equiv 1$).

In a tight-binding model of mobile electrons, the lattice distortion modifies the hopping amplitudes because wavefunction overlaps change. Empirically, the hopping amplitudes follow
\begin{equation}\label{nonlin-hoppings}
t_{ii'} = t_0 \exp\left[ - \beta(|\bs{\delta}_{ii'}|/a_0 -1)\right]
\end{equation}
where $t_0$ is the hopping in the absence of strain. $\bs{\delta}_{ii'}=\bs{r}_i+\bs{u}_i-\bs{r}_{i'}-\bs{u}_{i'}$ is the distance between sites $i$ and $i'$, and $\bs{u}$ is typically evaluated at the lattice positions $\bs{r}_i$. The factor $\beta$ encodes the strength of electron-lattice coupling; typical values are of order unity, for graphene\cite{neek-amal-13prb115428} $\beta=3.37$.
The dimensionless product $C = \bar{C}\beta$ (loosely referred to as ``strain'' below) measures the influence of the distortion on the hopping. For $C\ll 1$, one may linearize
\begin{equation}\label{lin-hoppings}
t_{ii'} = t_0 \left[ 1 - \beta(|\bs{\delta}_{ii'}|/a_0 -1)\right]\,.
\end{equation}

From a theoretical point of view, a given value of $C$ may be realized, on the one hand, with small $\beta$ and large $\bar{C}$: This is the case of weak electron-lattice coupling and implies a strongly distorted lattice. On the other hand, one may consider large $\beta$ and small $\bar{C}$, i.e., large electron-lattice coupling and weak distortions. 
Of particular interest is the limit $\beta\to\infty$ where the lattice distortion (for given $C$) is infinitesimal -- this is the limit of relevance for this paper. In this limit, the bond-length changes arise from longitudinal displacements only, reducing non-linearities in the hopping modulation.
Explicitly, in this limit we have $|\bs{\delta}_{ii'}| \approx a_0 + (\bs{u}_i-\bs{u}_{i'})\cdot {\dhat}_j$ for a bond along ${\dhat}_j$, \ie $\bs{r}_i-\bs{r}_{i'}/|\bs{r}_i-\bs{r}_{i'}| = {\dhat}_j$. The displacement difference can be approximated via the corresponding derivative, $\bs{u}_i-\bs{u}_{i'} \approx {\dhat}_j \cdot (\bs{\nabla} \bs{u}) a_0$; this is {\em exact} for the quadratic displacement fields of interest here. Eventually Eq.~\eqref{lin-hoppings} becomes
\begin{equation}\label{lin-hoppings2}
t_{ii'} = t_0 \left( 1 - \beta {\dhat}_j \cdot (\bs{\nabla} \bs{u}) \cdot {\dhat}_j \right)\,.
\end{equation}
For quadratic displacement fields, this yields a linear spatial variation of the hopping amplitudes. For the concrete cases of $\bs{u}^{\ }_{\rm 2D}$ and $\bs{u}^{\ }_{\rm 3D}$, combined with the 
${\dhat}_j$ vectors of the honeycomb and diamond lattice, respectively, one can check that Eq.~\eqref{lin-hoppings2} yields Eq.~{\eqhopping} of the main text, after a rescaling $t \to t(d+1)/N$ and choosing $C=\bar{C}\beta=d/(2N)$. This value of $C$ cuts out a subsystem of linear size $N$, or -- for given system size $N$ -- can be viewed as the maximum strain $C_{\rm max}(N)$ such that all hoppings remain positive.

\subsection{Triaxial strain on the honeycomb lattice}

Triaxial strain on the honeycomb lattice has been extensively discussed in the literature,\cite{PhysRevLett.101.226804,guinea-10np30,vozmediano-10pr109,levy-10s544,gomes-12n306} with the displacement field $\bs{u}_{2D}$ given above. 
For this case we now illustrate the evolution of the single-particle spectrum under triaxial strain as function of $\beta$, the strength of electron-lattice coupling. As argued above, in the limit $\beta\to\infty$ the hopping pattern resulting from triaxial strain smoothly transforms into the hopping modulation Eq.\,{\eqhopping} discussed in the main paper. As a result, the density of states (DOS) reveals that the approximate Landau level structure around zero energy changes into perfectly degenerate levels in the entire spectrum.
 
\begin{figure}[h!]
\centering
\includegraphics[scale=1.1]{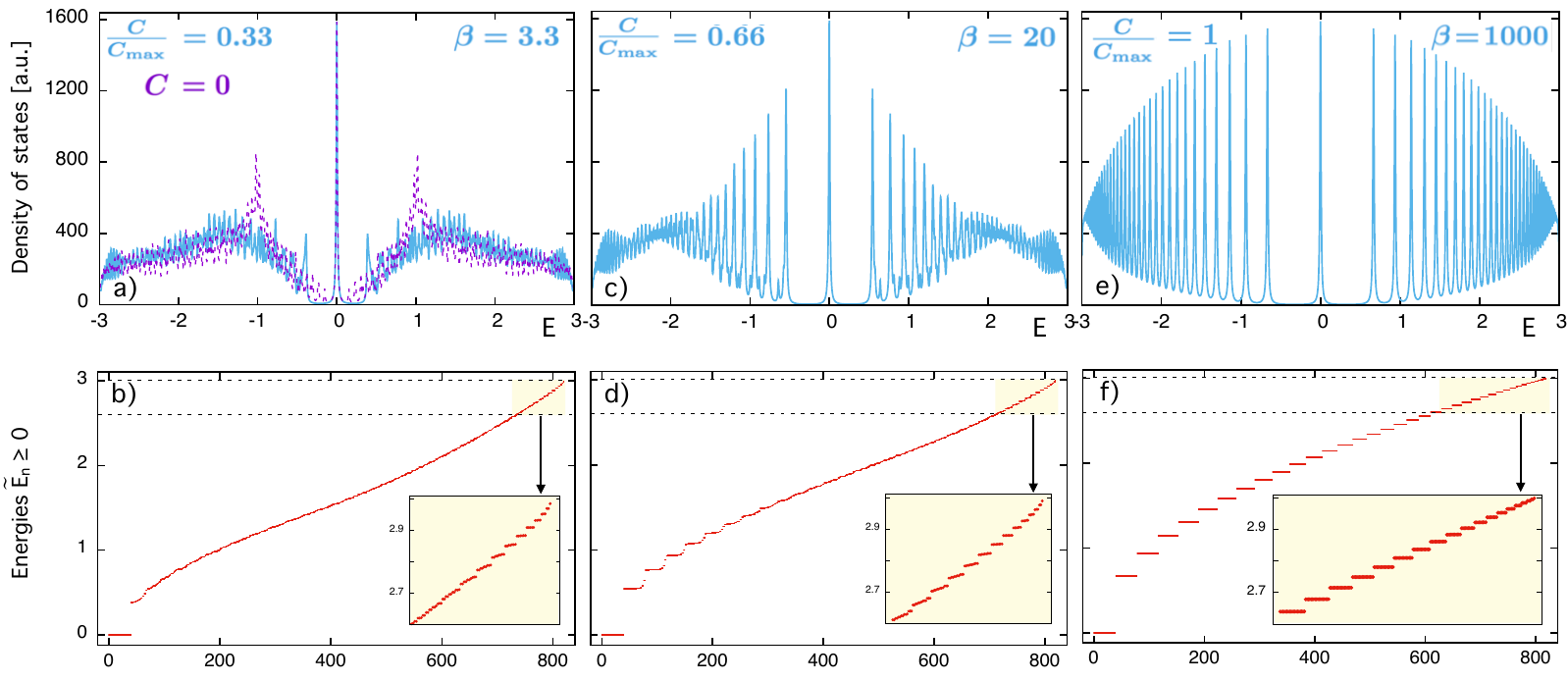}
\caption{DOS and eigenvalues for different strain values $C$ and electron-lattice couplings $\beta$ for a honeycomb-lattice triangle of size $N=40$ subject to triaxial strain.
Top row: DOS plots vs.\ energy $E$. Bottom row: Energy eigenvalues with ${\tilde{E}}_n \geq 0$ in increasing order; negative eigenvalues have been omitted (but follow from particle-hole symmetry). The high-energy range $2.6 \leq E \leq 3$ is shown in the inset.
Parameters used: (a, b) $C/C_{\rm max}=0.33$ and $\beta=3.3$ (panel (a) contains the unstrained DOS plot for comparison (dashed purple line)); (c, d) $C/C_{\rm max}=0.66$ and $\beta=20$; (e, f) $C/C_{\rm max}=1$ and $\beta=1000$.
$C_{\rm max}(N)$ refers to the maximum strain such that all hoppings remain positive.
The DOS plots employ a Lorentzian broadening of width $\gamma/t_0 = 0.008$.
}
\label{fig:figure2}
\end{figure}

Fig.\,\ref{fig:figure2} shows numerical results for a triangle of size $N=40$, corresponding to 1600 lattice sites, with hopping amplitudes calculated from Eq.~\eqref{lin-hoppings} with $t_0=1$. In panels (a) and (b), realistic $\beta$ values for graphene are chosen (in panel (a), also the unstrained ($C=0$) DOS is shown in purple for comparison). In panels (c) and (d), strong strain and $\beta=20$ already features quite pronounced Landau levels. Eventually, in panels (e) and (f) maximum strain and $\beta=1000$ (differences to $\beta=\infty$, \ie to Eq.\,{\eqhopping}, are beyond the resolution of these plots) is shown featuring perfect Landau levels following the prediction
\begin{equation}
{\tilde{E}}_n^\pm \equiv \frac{3}{N} E_n^\pm=\pm\frac{3}{N} \sqrt{N^2 - n^2}
\end{equation}
for $N=40$ and $n=1,\ldots, N$. Here, the rescaling factor $3/N$ as opposed to Eq.\,{\eqspec} has been implemented such that the spectrum is bounded between $+3$ and $-3$ allowing comparison with unstrained graphene.

\subsection{Tetraxial strain on the diamond lattice}

In the following, we consider tetraxial strain on the diamond lattice and aim to estimate how much strain is needed to create pseudo-Landau levels. Tetraxial strain is implemented by using hopping amplitudes \eqref{lin-hoppings} with the displacement field $\bs{u}_{\rm 3D}$ as defined in Sec.\,\ref{sec:VIA}.
For consistency, we still consider a diamond lattice having the shape of a tetrahedron. 
By analogy to triaxial strain on the honeycomb lattice we expect the  Landau level gap between the levels with $\tilde\eps_0=0$ and $\tilde\eps_1>0$ to form first; $\tilde\eps_n$ denotes the positive branch of energies for a diamond tight binding model and $\tilde\eps_0 = 0$ corresponds to the zero modes. We note that finite-size effects in $d=3$ are more serious than in $d=2$, such that it is difficult to distinguish finite-size gaps from Landau level gaps for small strain and accessible system sizes.
Therefore, we consider the lowest Landau level gap $\tilde\eps_1-\tilde\eps_0$  as a measure for the formation of Landau quantization;
in order to make the computations more efficient, we rather consider $H^2$ leading to energies ${\tilde\eps}^2$ (note that the system remains bipartite and all our findings for bipartite graphs carry over to the tetraxially strained diamond lattice). Consequently, we use $\tilde\eps_1^2-\tilde\eps_0^2$ as a measure for the lowest Landau level gap.

\begin{figure}[t!]
\centering
\includegraphics[scale=0.95]{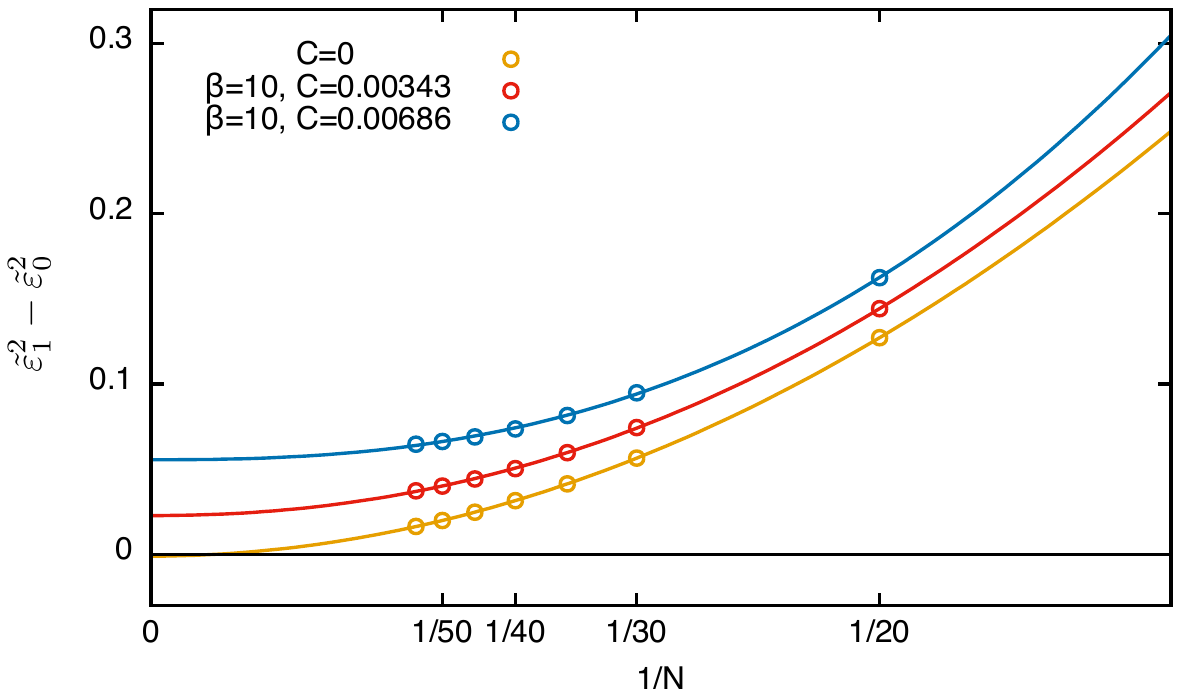}
\caption{Scaling of the energy gap between zero modes and the next higher energy for the diamond lattice in the absence of strain ($\bar C=0$, in yellow), and for two finite strain values ($\bar C = 0.000343$, $\beta=10$, in red; $\bar C = 0.000686$, $\beta=10$, in blue). These strain values correspond to $C/C_{\rm max}=0.05$ and $0.1$, respectively, for system size $N=50$.}
\label{fig:gap-tetraxial}
\end{figure}

For small system size $N$ and absence of any strain, there is an energy gap between the zero modes and the Dirac-type low-energy spectrum. Finite-size scaling clearly reveals that this is gap is due to finite size and disappears for $N\to\infty$ (yellow curve in Fig.\,\ref{fig:gap-tetraxial}). 
Now we  consider finite strain which we keep fixed while performing finite-size scaling, \ie we also fix the pseudo-magnetic field. We choose $\bar{C}=0.000343$ and $0.000686$ leading to $C/C_{\rm max}=0.05$ and $0.1$, respectively, for system size $N=50$ which corresponds to 42\,925 lattice sites. Already for 5\% strain we observe a clear energy gap (red curve in Fig.\,\ref{fig:gap-tetraxial}). This gap is rapidly increasing with increasing strain (blue curve); the behavior is consistent with $\tilde\eps^2 \propto \bar C$, as expected for Landau levels in Dirac-type systems.

These observations allow us to conclude that already moderate strain values of 5\% will clearly be sufficient to detect Landau quantization in 3D and that most of the previous findings for strained graphene are likely to be present for the strained diamond lattice as well. Indeed we find that tetraxial strain on the diamond lattice provides a promising path to realize 3D pseudo-Landau levels.

%

%
%
\section{Nature of the zero modes in the absence of strain}

As pointed out in the paper, an imbalance of A and B sites of a bipartite graph  gives rise to $|n_A - n_B|$ zero modes in the spectrum (this is the Lieb--Sutherland theorem). In the presence of the strain modulation {\eqhopping} we attributed these zero modes to the zeroth Landau level. In the following we give an physical interpretation of these zero modes when strain is absent, \ie for the corresponding homogenous tight-binding problem.

\subsection{$\bs{d=2}$}

\begin{figure}[h!]
\centering
\includegraphics[scale=0.6]{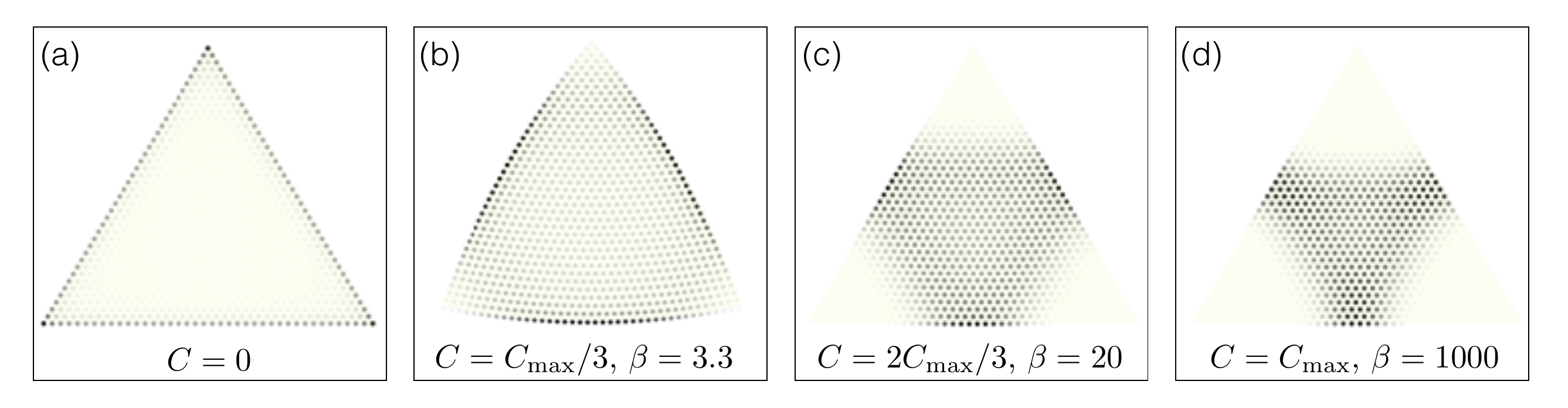}
\caption{
Integrated amplitude of all zero-energy eigenvectors for $d=2$ and size $N=40$; black (white) refers to large (vanishing) amplitude.
(a) Without strain $C=0$. (b) Moderate strain $C/C_{\rm max}=1/3$ with electron-lattice coupling $\beta=3.3$ as in graphene. (c) Intermediate strain $C/C_{\rm max}=2/3$ and stronger $\beta=20$. (d) Maximum strain $C=C_{\rm max}$ and limit of dominating $\beta=1000$.
The yellow triangle indicates the sample, here with the lattice deformation (clearly visible in panel (b)) included.
A  figure similar to panel (d) appears in Ref.~\onlinecite{poli-14prb155418}.
}
\label{fig:2d-zeromodes}
\end{figure}

In Fig.\,\ref{fig:2d-zeromodes} we show representative plots of all eigenfunctions at zero energy $E=0$. In the absence of strain, these zero modes correspond to the edge modes of graphene's zigzag edges. For finite strain, it turns out, however, that the zero modes are bulk modes and can be identified with the zeroth Landau level.

As an aside, we mention that a triangle with armchair edges (as opposed to zigzag edges considered throughout the paper) does not display zero modes in the absence of strain, because $n_A=n_B$ in this case. This is related to the fact that armchair edges of graphene are known not to exhibit any edge modes. Brief discussion of the interplay of armchair edges and strain are in Refs.~\onlinecite{poli-14prb155418} and \onlinecite{neek-amal-13prb115428}.

\subsection{$\bs{d=3}$}

\begin{figure}[h!]
\centering
\includegraphics[scale=0.75]{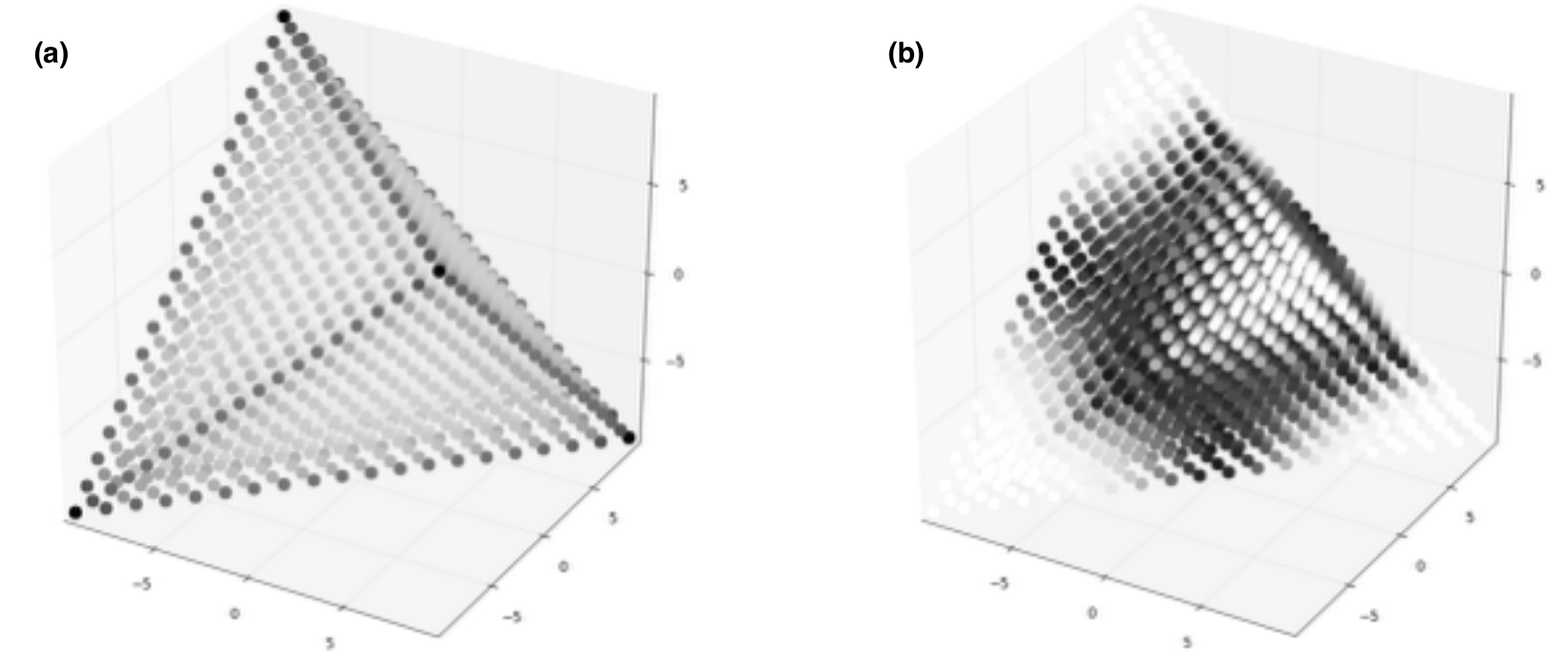}
\caption{
Same as Fig.~\ref{fig:2d-zeromodes}, but now for $d=3$ and $N=15$.
(a) Without strain $C=0$. (b) Maximum strain $C=C_{\rm max}$ and $\beta\to\infty$ corresponding to the hopping pattern in Eq.\,{\eqhopping}.
For the sake of clarity, all A sites have been omitted because the corresponding amplitudes are identically zero.
}
\label{fig:3d-zeromodes}
\end{figure}

For $d=3$, we find a similar situation: in the absence of strain, the zero modes correspond to states wich are localized at the edges of the tetrahedron. For maximal strain, the zero modes correspond to bulk states which appear as a natural generalization of the $d=2$ case shown in Fig.\,\ref{fig:2d-zeromodes}\,(d); this further substantiates the interpretation as three-dimensional Landau levels.

\bibliographystyle{prsty}
\bibliography{triangles}